\documentclass[prl,twocolumn]{revtex4}

\usepackage{graphicx}
\usepackage{amsmath}

\begin{document}

\title{Dark--Bright Solitons in Inhomogeneous Bose-Einstein Condensates}

\author{Th.~Busch$^1$ and J.~R.~Anglin$^2$}

\affiliation{$^1$Institute of Physics and Astronomy,
                 Aarhus University,
                 Ny Munkegade,
                 DK--8000 \AA rhus C, Denmark\\
             $^2$ITAMP Harvard-Smithsonian CfA,
                 60 Garden Street, MS 14,
                 Cambridge, MA 02138}

\date{\today}

\begin{abstract}
  We investigate dark--bright vector solitary wave solutions
  to the coupled non--linear Schr\"odinger equations which describe an
  inhomogeneous two-species Bose-Einstein condensate.  While these
  structures are well known in non--linear fiber optics, we show that
  spatial inhomogeneity strongly affects their motion, stability, and
  interaction, and that current technology suffices for their creation
  and control in ultracold trapped gases. The effects of controllably
  different interparticle scattering lengths, and stability against
  three-dimensional deformations, are also examined.
\end{abstract}


\maketitle

% Introductory paragraph

Among the many surprising features of non--linear equations, the
emergence of solitons is one of the most prominent. For the
non--linear Schr\"{o}dinger equation (NLSE), which governs both
non--linear optical modes in fibers and dilute Bose-Einstein
condensates (BECs), two different kinds of scalar solitons, {\it 
  bright} and {\it dark}, are known \cite {Shabat}. In optics, bright
and dark solitons arise in media with anomalous dispersion and normal
dispersion, respectively, and for BECs the s--wave scattering
interaction is the determining factor (attractive for bright solitons,
repulsive for dark). Whereas in gaseous Bose--condensates dark
solitons only, and only recently, have been observed
\cite{Hannover,Nist}, optical solitons are well investigated and on
the verge of industrial application \cite{SolitonReviews}. In addition
to the bright and dark scalar solitons there are also various
multicomponent (vector) solitons known, which arise as solutions to
systems of coupled NLSEs. An elegant example is the so--called {\sl
  dark--bright} soliton, where a bright optical solitary wave exists
in a system with normal dispersion because it is trapped within a
copropagating dark soliton \cite {Trillo,Christodoulides,Shalaby}. In
this Letter we investigate the behavior of dark--bright solitons (and
solitary waves) in repulsively interacting two-component Bose-Einstein
condensates. We examine the effects of spatial inhomogeneity, three
dimensional geometry, and dissipation, which are all important
features of BEC experiments.

% Bright-dark solitons in bulk

In the context of cold atomic gases the two vector components evolving
under the Gross-Pitaevskii NLSE are the macroscopic wave functions of
Bose-condensed atoms in two different internal states, which we will
denote as $|D\rangle $ and $|B\rangle $. The non--linear interactions
are due to elastic s--wave scattering among the atoms, and are
effectively repulsive (positive scattering length) for both systems
($^{23}$Na and $^{87}$Rb) in which multicomponent condensates have
been realized. In both cases the three 1D interaction strengths
($g_{DD},g_{BB},g_{DB}$) may easily be made equal to within a few
percent; and quasi-one dimensional traps that are longitudinally very
flat are under active experimental development \cite{1DTraps}. So we
will begin with the two-component NLSE with all interaction strengths
equal, in one dimension, with no potential, and then add more
realistic features in succession. In the standard natural units, the
general equations read
\begin{eqnarray}
i\dot{\psi}_{B} &=&-\frac{1}{2}\psi _{B}^{\prime \prime }+[V_{B}+|\psi
_{B}|^{2}+G_{D}|\psi _{D}|^{2}-\mu -\Delta ]\psi _{B}\;,  \nonumber \\
i\dot{\psi}_{D} &=&-\frac{1}{2}\psi _{D}^{\prime \prime }+[V_{D}+|\psi
_{D}|^{2}+G_{B}|\psi _{B}|^{2}-\mu ]\psi _{D}\;,  
  \label{eq:GPEs}
\end{eqnarray}
where the chemical potentials $\mu _{D}=\mu $ and $\mu _{B}\equiv \mu
+\Delta $ have been introduced in the standard way. For the present we set
the coupling strength ratios $G_{B,D}=1$ and assume no external potentials, $%
V_{D,B}=0$. The dark--bright soliton solution to eqs.~(\ref{eq:GPEs}) is
then given by 
\begin{eqnarray}
\psi _{B} &=&\sqrt{\frac{N_{B}\kappa }{2}}e^{i\phi }e^{i\Omega
_{B}t}e^{ix\kappa \tan \alpha }\hspace{2pt}\hspace{2pt}\text{sech}\Bigl%
(\kappa \lbrack x-q(t)]\Bigr)\;,    \nonumber \\
\psi _{D} &=&i\sqrt{\mu }\sin \alpha +\sqrt{\mu }\cos \alpha \tanh \Bigl%
(\kappa \lbrack x-q(t)]\Bigr)\;,
  \label{eq:BiDS}
\end{eqnarray}
where $N_{B}\equiv \int \!dx\,|\psi _{B}|^{2}$ is the rescaled number
of particles in state $|B\rangle $, the soliton inverse length is
$\kappa \equiv \sqrt{\mu \cos ^{2}\alpha +(N_{B}/4)^{2}}-N_{B}/4$, the
bright component frequency shift is $\Omega _{B}\equiv \kappa
^{2}(1-\tan ^{2}\alpha )/2-\Delta $, and the soliton position is
$q(t)=q(0)+t\kappa \tan \alpha $. The `binding energy' of the bright
component in the well formed by the $\psi _{D}$ mean field, in the
co-moving frame, is clearly $\kappa ^{2}/2 $; the bright component
phase shift $\phi $ is only of significance if there are two or more
solitons.
\begin{figure}[tbp]
  \centering
  \includegraphics[width=0.95\linewidth,clip=true]{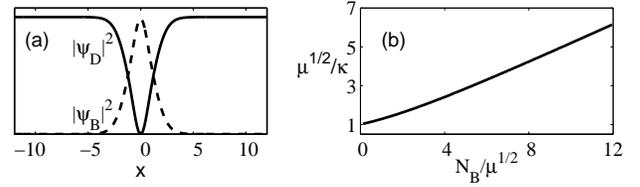}
  \caption{$(a)$ A dark--bright soliton solution of eqs.~(\ref{eq:BiDS}), with 
    $\protect\alpha =0$. The rescaled densities of the bright and dark
    components, $|\protect\psi _{B}|^{2}$ and $|\protect\psi
    _{D}|^{2}$, are shown with a broken and a full line, respectively.
    $(b)$ The size of a motionless dark--bright soliton, in units of
    the healing length $\protect\mu^{-1/2}$, as a function of
    $N_{B}\protect\mu ^{-1/2}$.}
\label{fig:FDSoliton}
\end{figure}
Readers familiar with the scalar solitons of the one-component NLSE
will recognize $\psi_D$ as a dark (or `grey') soliton of
velocity-angle $\alpha$, and $\psi_B$ as a bright soliton, which can
only be found in single-component condensates if they have negative
scattering length (and so are prone to collapse) (see
Fig.~\ref{fig:FDSoliton}(a)). The Thomas-Fermi-like expansion with
$N_B$ of the trapped bright component makes the soliton size
$\kappa^{-1}$ longer than for a single-component dark soliton at the
same $\mu$ (see Fig.~\ref{fig:FDSoliton}(b)).

% Equation of motion in 1D traps

The integrable system of eqs.~(\ref{eq:GPEs}) with $G_{D,B}=1$ and
$V_{D,B}=0 $, also known as the Manakov equation, conserves the free
energy
\begin{align}
 G&=\frac{1}{2}\int \!dx\Bigl[|\psi _{D}^{\prime }|^{2}+|\psi _{B}^{\prime
}|^{2}+(|\psi _{D}|^{2}+|\psi _{B}|^{2}-\mu )^{2} \nonumber\\
  &\hspace{6cm}+2\Delta |\psi_{B}|^{2}\Bigr]  \nonumber \\
&=\frac{4}{3}\kappa ^{3}+{\frac{1}{2}}N_{B}\kappa ^{2}(1+\tan ^{2}\alpha
)+N_{B}\Delta \;.  \label{eq:Gbulk}
\end{align}
Since $G$ decreases with increasing soliton velocity, the soliton is
formally unstable (to acceleration!). But one implication of
integrability is that perturbations of eqs.~(\ref{eq:BiDS}) due to
interactions with other waves (solitary or ordinary) will not cause
dissipation. If an inhomogeneous potential is added, however, by
allowing non--zero $V_{D,B}(x)$ in (\ref {eq:GPEs}), then the system
is no longer integrable, and the soliton can interact non-trivially
with the surrounding condensate. Nevertheless, if $V $ varies slowly
on the soliton scale $\kappa $, then excitations of the background
that have high enough temporal frequency to accept energy resonantly
from the soliton must also have short enough spatial wavelength to
`see' $V_{D,B}(x)$ as approximately constant, and hence tend to
decouple from the soliton as they do in the truly integrable case. The
result is that $G(q,\dot{q})$ (given by replacing $\mu \rightarrow
\mu-V_{D}(q),\Delta \rightarrow \Delta -V_{B}(q)+V_{D}(q)$ in
eq.~(\ref{eq:Gbulk}) and inverting $\dot{q}=\kappa (q,\cos \alpha
)\tan \alpha $) is approximately conserved, and this determines the
motion of the soliton in the potential. (A more sophisticated multiple
time scale boundary layer analysis supports this simple argument.)

\begin{figure}[tbp]
  \centering
  \includegraphics[width=0.9\linewidth,clip=true]{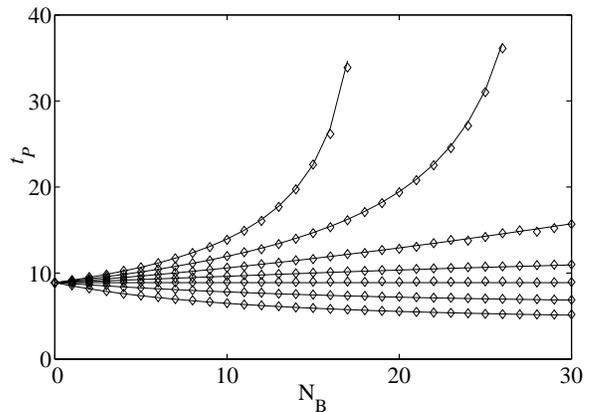}
  \caption{Period of oscillation for a dark--bright soliton in a harmonic trap
    calculated numerically (diamonds) and from eq.~(\ref{eq:EOMFDS})
    (solid line). The trapping potentials for the different components
    are related as $V_{B}=\protect\gamma V_{D}$, with
    $\protect\gamma=-1,-0.5,0.5,0.75,1,1.25,1.5$ respectively for the
    graphs starting from below. The normalization of the dark
    component is $\int dx|\protect\psi _{D}|^{2}=1000$ in all cases,
    so that $\protect\mu $ ranges from 66.2 to 67.1. The divergence of
    the curves for $\protect\gamma =1.25$ and $\protect\gamma =1.5$
    shows the breakdown of the oscillations as explained in the text.}
\label{fig:OscAmp}
\end{figure}

The motion this implies simplifies in the limit of velocities much
smaller than the speed of sound, because
\begin{eqnarray}
G &=&{\frac{4}{3}}\Bigl[\mu +{\frac{N_{B}^{2}}{16}}-V_{D}(q)\Bigr %
]^{3/2}+N_{B}\Bigl[V_{B}(q)-{\frac{V_{D}(q)}{2}}\Bigr]  \nonumber \\
&&-2\dot{q}^{2}\sqrt{\mu +{\frac{N_{B}^{2}}{16}}-V_{D}(q)}\;\;+{\cal O}(\dot{
q}^{4})\;,  \label{eq:Glow}
\end{eqnarray}
dropping a term which is constant if we assume that none of the
conserved state $|B\rangle $ atoms escape from the soliton. This
implies the low-velocity equation of motion
\begin{equation}
\ddot{q}=-{\frac{V_{D}^{\prime }(q)}{2}}-{\frac{N_{B}[V_{D}^{\prime
}(q)-2V_{B}^{\prime }(q)]}{8\sqrt{\mu +{N_{B}^{2}/16}-V_{D}(q)}}}\;,
\label{eq:EOMFDS}
\end{equation}
which, together with its numerical confirmation shown in Fig.~\ref
{fig:OscAmp}, is the primary result of this paper. In the limit
$N_{B}\rightarrow 0$ we recover the equation of motion of the dark
soliton \cite{BuschAnglin:2000}, and as $N_{B}$ increases we find that
the soliton is more and more insulated from the effect of $V_{D}$, and
more sensitive to $V_{B}-V_{D}$. In the limit $N_{B}\gg \sqrt{\mu }$,
where the soliton is expanded by the large bright component to many
healing lengths in size, we have
\begin{equation}
\ddot{q}= \left( 1-8{\frac{\mu -V_{D}}{N_{B}^{2}}}\right)
[V_{B}^\prime(q)-V^\prime_{D}(q)] -4{\frac{\mu -V_{D}}{N_{B}^{2}}}%
V^\prime_{D}(q)\;,
\end{equation}
so that a small differential force on the bright component will
predominate.  Our assumption that the whole soliton is small compared
to the trap scale, however, means that the dark component retains its
dramatic effect of giving the soliton an effectively negative mass:
the soliton accelerates in the opposite direction to a force exerted
through $V_{B}$. If $V_{B}$ and $V_{D}$ are kept equal, on the other
hand, a highly expanded dark--bright soliton with $N_{B}\gg
\sqrt{\mu}$ moves in the potential as if it had a very large positive
mass (because as one can see from eq.~(\ref{eq:Glow}) the soliton's
potential energy is also $\sim -V_{D}$). Numerical integration of the
coupled NLSEs shows excellent agreement with eq.~(\ref{eq:EOMFDS})
(see Fig.~\ref{fig:OscAmp}). Note that for harmonic $V_{B}>V_{D}$
eq.~(\ref{eq:EOMFDS}) implies that there is a critical $N_{B}$ above
which the soliton will escape from the trap instead of oscillating.
While the precise transition point between very slow oscillation and
very slow escape is difficult to check numerically, our numerical
results confirm that escape does occur in this case at larger $N_{B}$.

\begin{figure}[tbp]
  \centering
  \includegraphics[width=0.95\linewidth,clip=true]{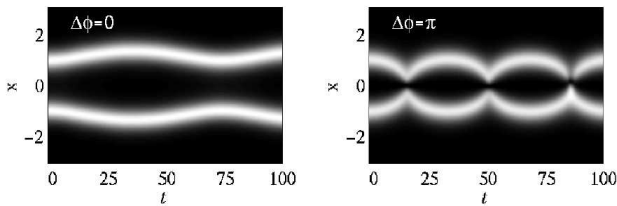}
  \caption{Symmetric collision of two dark--bright solitons in a harmonic trap
    ($\protect\gamma =1$). Degree of brightness indicates
    $|\protect\psi _{B}|^{2}$ as a function of $x $ and $t$ (both in
    trap units), for repulsive ($\Delta \protect\phi =0$) and
    attractive ($\Delta \protect\phi=\protect\pi $) interaction.}
\label{fig:TOsc}
\end{figure}

\begin{figure}[tbp]
  \centering
  \includegraphics[width=0.95\linewidth,clip=true]{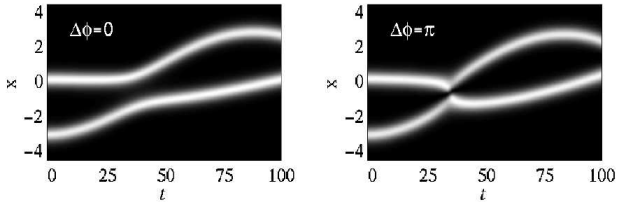}
  \caption{Collision of two dark--bright solitons in a harmonic trap
    ($\protect\gamma =1$). Shown is $|\protect\psi _{B}|^{2}$ as a
    function of $x$ and $t$, both in trap units.
    %In the plot on the left the two bright components
    %initially have the same phase, whereas in the plot on the right
    %they initially have opposite phase.  
    Initially one soliton is at rest (i.e.~$q=0$); after the
    collision both solitons are oscillating.}
\label{fig:Disp}
\end{figure}

A trapping potential also modifies the interactions between solitons.
While generally solitons that are more than a few soliton lengths
apart are essentially unaffected by each other, at closer ranges they
can distort each other significantly. Although the respective bright
component numbers $\propto N_{B}$ of two solitons are simply conserved
during such interactions, the relative phase of the two bright
components strongly affects the details of the interaction
\cite{Segev,Kivshar}: bright--dark solitons repel each other when the
phase difference between the bright components is $\Delta \phi =\phi
_{1}-\phi _{2}=0$, and attract each other when this difference is
$\Delta \phi =\pi $.  This short range behaviour, which is opposite to
that of scalar bright solitons, occurs independently of the confining
potential, but if the effect of a potential is to keep two solitons
close together, then their phase-dependent interaction can
significantly affect their oscillations: see Fig.~(\ref{fig:TOsc}).
Finally, even if the trap does not confine both solitons within their
interaction range, the inhomogeneous potential will still
qualitatively modify the effects of soliton collisions. In the
integrable case with no potential, soliton collisions are trivial, in
the sense that the asymptotic states that emerge after a collision are
the same as they were before it. In general, however, there is a net
effect of even an integrable collision: a `jump-like' spatial
translation of the solitons, relative to where each would have been if
it had not encountered the other.  Fig.~\ref{fig:Disp} shows that in a
trap such a translation can imply a transfer of energy between
solitons as a result of a collision.

In addition to the inhomogeneous potential, integrability may also be
destroyed under experimental conditions by the fact that the three
interaction strengths $g_{DD}, g_{BB}$ and $g_{DB}$ governing the
non--linearity will generally differ. In the quasi--1D limit, the
interaction strengths are given by $g_{ij}=a_{ij}/(A_{i}+A_{j})$,
where $a_{ij}$ is the 3D s-wave scattering length and $A_{j}$ is the
cross-sectional area of the trap confining species $j=D,B$. Since in
our standard natural units we have absorbed in $\psi _{D,B}$ the
self-interaction strengths $g_{DD}$ and $g_{BB} $, in this more
general case we must allow the co-efficients $G_{B}\equiv
g_{DB}/g_{BB}$ and $G_{D}\equiv g_{DB}/g_{DD}$ to differ from unity in
the NLSE system (\ref{eq:GPEs}). It is straightforward to show that
\begin{eqnarray}
G_{D} &=&\frac{a_{DB}}{a_{DD}}\left( 1+\frac{A_{D}-A_{B}}{A_{D}+A_{B}}%
\right) \;,  \nonumber \\
G_{B} &=&\frac{a_{DB}}{a_{BB}}\left( 1-\frac{A_{D}-A_{B}}{A_{D}+A_{B}}%
\right) \;,
\end{eqnarray}
so that varying the relative tightness of radial confinement for the
two species yields one free control parameter, which can allow
significant retuning of $G_{D,B}$ even without the measure of
modifying the scattering lengths themselves.

\begin{figure}[tbp]
  \centering
  \includegraphics[width=0.95\linewidth,clip=true]{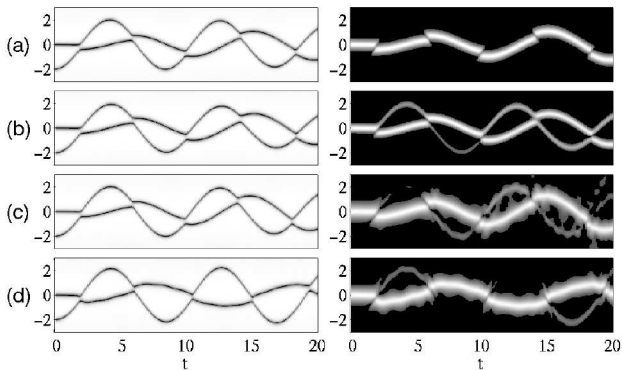}
  \caption{Collisions between an initially empty ($N_{B}=0$) dark soliton and
    a dark-bright solitary wave initially at rest in a harmonic trap.
    Each row is a separate evolution, with the left and right plots in
    each row showing $| \protect\psi _{D}|^{2}$ and $|\protect\psi
    _{B}|^{2}$ respectively.  Horizontal axes are time, vertical axes
    space, both in trap units. In all cases $\int \!dx|\protect\psi
    _{D}|^{2}=400$ and $\int \!dx|\protect\psi _{B}|^{2}=4$, for
    $\protect\mu \doteq 36$ and Thomas-Fermi radius $\doteq 8.6 $. For
    each row, from top to bottom, we have $(G_{D},G_{B})$ as follows:
    (a): (1.03,1.03); this case is not distinguishable from (1,1).
    (b): (1.5,1.5); transfer of bright component occurs. (c):
    (0.5,0.5); scattering as well as transfer of bright component is
    seen in the last collision. (d): (0.5,1.5); viewing a true `movie'
    of $|\protect\psi _{D}|^{2}$ reveals that, during the collisions,
    the background cloud is much more significantly excited in this
    case than in the others; and this is the reason for the noticeably
    different soliton motion in this case. The case (1.5,0.5), which
    is not shown here, is just noticeably different from (1.5,1.5).}
\label{fig:USymScatt}
\end{figure}

If $G_{D,B}\not=1$, solutions that are distorted versions of the
dark--bright soliton certainly exist, although they may only be given
in closed form for special cases. (For instance, in the limit of small
$N_{B}$ one may discard the $|\psi _{B}|^{2}$ terms in the NLSE, and
find dark soliton solutions for $\psi _{D}$, with $\psi _{B}\propto
\hspace{2pt}\hspace{2pt}\text{sech}^{\nu }[\kappa (x-q)]$ for $\nu
(\nu +1)=2G_{D}$. For $G_{D}>1$, there are also one or more excited
bound states of $\psi _{B}$.)  Such solutions are often referred to as
solitary waves, rather than solitons, to indicate that they may
interact nontrivially with other solitary or ordinary waves. This
means, for example, that a collision between dark--bright solitary
waves may effect a net transfer of bright component from one solitary
wave to the other; see Fig.~(\ref{fig:USymScatt}). It also implies
that unlike true solitons, which are transparent to all quasiparticle
modes, dark--bright solitary waves will suffer from dissipation due to
collisions with thermal particles and phonons, even when the
one-dimensional approximation is excellent. It is a problem beyond the
scope of this Letter to compute scattering rates with $G_{D,B}$
significantly different from unity. For $G_{D,B}$ close to unity,
however, simple estimates for the anti-damping rate due this effect
show it to be negligible, at attainably low temperatures, because the
cross sections for dissipative collisions are proportional to the
squares of the scattering length differences.
In current experiments, however, soliton lifetimes are limited not by
one-dimensional dissipation, but by the breakdown of the
one-dimensional approximation. In more than one dimension,
single-component dark solitons suffer from the well-known `snake mode'
dynamical instability, in which bending of the plane of the density
minimum grows exponentially even without dissipation. This is the
presumed cause of decay of solitons that have been produced in BEC
experiments to date. As shown by Muryshev {\it et al.}
\cite{Shlyapnikov}, a single-component dark soliton is only unstable
to snake modes of wavelength greater than the soliton size, so that
radial confinement to within a healing length should stabilize dark
solitons. By an extension of their method of analysis, one can show
that the dark--bright soliton is also stable against snake modes of
wavelength less than its size $\kappa ^{-1}$. Since for $N_{B}\gg
\sqrt{\mu }$ this can be much larger than the healing length,
stabilization of dark--bright solitons against snaking should easily
be possible in current traps. (The proof is simple in the case of a
motionless dark-bright soliton in three-dimensional bulk. We take
eqs.~(\ref{eq:BiDS}) for $\alpha =0$, and expand the free energy $G$
to second order in $\delta \psi _{j}=\Phi _{j}(x)\cos k(y\cos \theta
+z\sin \theta -\beta )$, where $R_{j}$ and $S_{j}$ are real, and
$j=B,D.$ The result for this perturbation of wave number $k$ is
\begin{eqnarray}
\delta G_{k} &=&\int dx\,\Gamma _{k}(x)\int dy\,dz\,\cos ^{2}k(y\cos \theta
+z\sin \theta -\beta )  \nonumber \\
\Gamma _{k}(x) &\equiv &2\left[ \sqrt{\mu }\Re \lbrack \Phi _{D}]\kappa
\tanh (\kappa x)+\sqrt{\frac{N_{B}\kappa }{2}}\Re \lbrack \Phi _{B}]\hspace{%
2pt}\text{sech}(\kappa x)\right] ^{2}  \nonumber \\
&&+\Phi _{D}^{\ast }\hat{H}_{k}\Phi _{D}+\Phi _{B}^{\ast }\left( \hat{H}_{k}+%
\frac{\kappa ^{2}}{2}\right) \Phi _{B}  \nonumber \\
\hat{H}_{k} &=&-\frac{1}{2}\frac{d^{2}}{dx^{2}}+\frac{k^{2}}{2}-\kappa ^{2\,}%
\hspace{2pt}\text{sech}^{2}(\kappa x)\;.
\end{eqnarray}
Hence, for the $k$ sector to have positive definite free energy, it is
obviously a sufficient condition that $\hat{H}_{k}$ have a positive
spectrum. Considering the possibility of purely imaginary $\Phi _{D}$
shows that this is also a necessary condition. \ Considering the
eigenfunction sech\thinspace $\kappa x$ (which is clearly the ground
state because it is real and nodeless) shows that the spectrum of this
particular $\hat{H}_{k}$ is positive for $|k|>\kappa $.)

% Creation

Even more conveniently, we note that a robust method for the
controlled creation of dark--bright solitons has already been
presented and analysed in detail (without being explicitly recognized
as such) \cite{Dum}. Dum {\it et al.} have shown that dark solitons
may be created in one condensate component by adiabatic transfer of
population from a condensate in another internal state, and have
already noted that in the late stages of this procedure the second
component appears as a stretched dark soliton around the remaining
population of the first, whose wave function approaches a hyperbolic
secant. Stopping short of complete adiabatic transfer will in fact
produce a dark-bright soliton of arbitrary $N_B$. Engineering by
masked Rabi transfer with phase imprinting may also be possible, and
the smoother total density profile and larger size of a
highly-expanded dark--bright soliton should simplify the creation of
slow and stable solitons by this method.

Our conclusion therefore is that dark--bright solitons move in trapped
condensates much as dark solitons (though more slowly, if $V_D=V_B$),
and have strong advantages in stability and controllability. Since
$|\psi_B|^2$ can be imaged separately and non-destructively, they also
offer a realization of Reinhardt's and Clark's proposal to track
solitons by trapping distinguishable atoms inside them\cite{Clark}.
And in addition to advancing soliton studies into the inhomogeneous
regime of BECs, production of dark-bright solitons in BECs would be
the development of atom optical tweezers with potentially sub-micron
precision: the trapping and manipulation of ultracold atoms by
ultracold atoms.

\begin{acknowledgments}
  We gratefully acknowledge discussions with J.I.~Cirac, P.~Zoller,
  Q--H.~Park and K.~M\o lmer, and support by the European Union (under
  the TMR network No.~ERBFMRX-CT960002), the Austrian FWF, the Danish
  research council, and the American NSF (through its grant to the
  Institute for Theoretical Atomic and Molecular Physics at the
  Harvard-Smithsonian Center for Astrophysics).
\end{acknowledgments}

% --------------------------------------------------------

\end{document}